\documentclass[%
 reprint,
 amsmath,amssymb,
 aps,
nofootinbib,
]{revtex4-1}

\usepackage{graphicx}% Include figure files
\usepackage{dcolumn}% Align table columns on decimal point
\usepackage{bm}% bold math

\begin{document}

\title{Thermodynamics of spacetime from minimal area}

\author{A. Alonso-Serrano}
\email{ana.alonso.serrano@aei.mpg.de}
\affiliation{Max-Planck-Institut  f\"ur  Gravitationsphysik  (Albert-Einstein-Institut),\\Am M\"{u}hlenberg 1, 14476 Potsdam, Germany}
	
\author{M. Li\v{s}ka}
\email{liska.mk@seznam.cz}
\affiliation{Institute of Theoretical Physics, Faculty of Mathematics and Physics, Charles University,
V Hole\v{s}ovi\v{c}k\'{a}ch 2, 180 00 Prague 8, Czech Republic}
\affiliation{Max-Planck-Institut  f\"ur  Gravitationsphysik  (Albert-Einstein-Institut), \\
Am M\"{u}hlenberg 1, 14476 Potsdam, Germany}

\begin{abstract}
Motivated by exploring the interface between thermodynamics of spacetime and quantum gravity effects, we develop a heuristic derivation of Hawking temperature and Bekenstein entropy from the existence of a minimal resolvable area. Moreover, we find leading order quantum gravity corrections to them that are in qualitative agreement with results obtained by other methods, both heuristic and rigorous. In this way, we recover, as a particular case, the corrections heuristically obtained from the existence of minimal length. We also show that the size of minimal area is constrained from above by well-understood results of semiclassical black hole physics, specifically by the entropy content of Hawking radiation. The minimal area derivation we introduce is also applied to finding the Unruh temperature associated with causal diamonds and to establish a new relation between this temperature and the entropy of the causal diamond's horizon.

\end{abstract}
	
\maketitle

\section{Introduction}

In the study of gravitational phenomena, the introduction of thermodynamic tools has revealed a useful approach to understand different processes involving horizons, such as black hole evaporation. Black hole thermodynamics emerged in the early 1970s with the argument that black holes posses entropy proportional to their horizon area~\cite{Bekenstein:1973}. Soon thereafter, it was found that black hole evolution follows four laws analogous to the laws of thermodynamics~\cite{Bardeen:1973} and that they emit black body radiation corresponding to a finite temperature~\cite{Hawking:1975} (the Hawking effect). The ideas behind the Hawking effect were later further extended and it was found that a uniformly accelerating observer perceives the Minkowski vacuum as a thermal bath of particles whose temperature is proportional to the observer's acceleration~\cite{Unruh:1976} (while this phenomena, known as the Unruh effect, is related to the Hawking effect through the equivalence principle, they are nevertheless distinct~\cite{Barbado:2016}).

In order to get a better understanding of these effects and develop new predictions, many results concerning the relation of thermodynamics and gravity have been later reproduced and extended by rather simple heuristic arguments. For instance, expressions of Hawking and Unruh temperatures are implied by the uncertainty relation between position and momentum of the produced particles~\cite{Scardigli:1995}. The form of Bekenstein entropy then follows from the Hawking temperature and the equilibrium Clausius relation, providing a complete thermodynamic description of a Schwarzschild black hole just from heuristic considerations. While these derivations of course cannot replace the corresponding rigorous calculations, they still provide useful physical intuition and allow one to develop new predictions and directions to explore meticulously in the future. Furthermore, one can extend heuristic reasoning beyond the well-explored semiclassical setting and employ it to gain insight into low energy quantum gravity effects. To achieve this, the standard Heisenberg uncertainty principle has been replaced by the generalised uncertainty principle (GUP)~\cite{Adler:2001,Alonso:2018a,Scardigli:2018jlm,Alonso:2021}, that phenomenologically incorporates minimal resolvable length appearing in a number of approaches to quantum gravity~\cite{Garay:1994en,Hossenfelder:2013}.

In this paper, inspired by previous successes of heuristic results, we park for a while the more rigorous side of black hole thermodynamics and drive our ideas to explore new directions and results. The nature of our reasoning is similar to that of the derivations based on GUP. However, in place of minimal length we introduce a new game piece: minimal resolvable area. There are good reasons to consider it. While the existence of minimal length trivially implies minimal area, the converse does not hold. To see this, consider an ellipse with height $2b$ and width $2a$, whose area, $\mathcal{A}=\pi ab$ corresponds to the minimal value. One can then set the height arbitrarily small, as long as its product with width and, thus, the area, remains fixed. Moreover, some approaches to quantum gravity say something about minimal area but not necessarily about minimal length, e.g. loop quantum gravity (LQG)~\cite{Rovelli:1995,Ashtekar:2021} and proposals to quantise the area of a black hole horizon~\cite{Bekenstein:1995}.

We begin by finding a derivation for the modified black hole temperature and entropy from minimal area. As expected, whenever the GUP approach is applicable, we obtain results equivalent to it. However, as we emphasised above, our findings are more general and hold even for some theories that do not necessarily imply GUP. Upon verifying and generalising the known previous results, we proceed to ask two new questions. First, if minimal area allows us to recover some predictions concerning black hole thermodynamics, can we in turn use our knowledge of black hole thermodynamics to learn something about minimal area? We argue that average entropy per photon of Hawking radiation indeed provides an upper limit on the minimal area and, indirectly, the minimal length. Since the semiclassical entropy of Hawking radiation depends only on its black body nature~\cite{Alonso:2018b}, the limits obtained in this way are model independent. Furthermore, the upper bound we find for the minimal area, $\mathcal{A}_{min}\lesssim10.80l_P^2$, is consistent with theoretical predictions of its value~\cite{Bekenstein:1995,Ashtekar:2006}. The second question we pose comes from the fact that one can assign entropy proportional to area even to observer-dependent horizons associated with accelerating observers~\cite{Jacobson:2003,Solodukhin:2010,Jacobson:2015}, who measure non-zero Unruh temperature. Therefore, it is natural to ask: can one connect the values of horizon entropy and Unruh temperature associated to such horizons? Applying our derivation based on minimal area, we show that this is indeed possible in the case of causal diamonds (and, possibly, other closed horizons). It turns out that the relation between the Unruh temperature measured by finite lifetime inertial observers inside the diamond~\cite{Martinetti:2003,Arzano:2020} and the entropy of diamond's horizon is similar as in the case of temperature and entropy of a Schwarzschild black hole. Furthermore, we derive corrections to diamond's temperature and entropy, obtaining results consistent with the GUP modified Unruh temperature~\cite{Scardigli:2018jlm} and the entanglement entropy of a spherical horizon in Minkowski spacetime~\cite{Solodukhin:2010}.

While we do not aim to provide a solid closed answer to any of these questions, we offer a novel and intuitive viewpoint that might prove helpful in solving them completely and even motivate further research in related topics. In any case, a slightly playful approach we adapt seems fitting for a field of research that is said to begin with a question: "What happens when you pour a cup of tea into a black hole\footnote{Reportedly, the question was posed by J.~A.~Wheeler to his student, J.~D.~Bekenstein, who consequently proposed his formula for black hole entropy~\cite{Wheeler:1998}.}?"

The paper is organised as follows. In section~\ref{Schwarzschild} we derive the modified temperature and entropy of a Schwarzschild black hole and discuss bounds on minimal area implied by black hole evaporation. Section~\ref{GLCD} is devoted to derivation of temperature and entropy of a causal diamond. We also explore how are both quantities modified due to quantum gravitational effects. Section~\ref{discussion} sums up our results and discusses unresolved issues.

Throughout the paper we will work in four spacetime dimensions, assume metric signature $(-1,1,1,1)$ and use SI units. Other conventions follow~\cite{MTW}.

\section{Thermodynamics of Schwarzschild black holes}
\label{Schwarzschild}

Our first scenario will be a Schwarzschild black hole. It makes for a convenient choice as it is fully described by a single parameter, the mass $M$. Therefore, we have a very simple dependence of entropy on temperature given by the equilibrium Clausius relation, $\text{d}S=c^2\text{d}M/T$. Later on, when we turn our attention to causal diamonds (section~\ref{GLCD}), we will see how the presence of thermodynamic quantities beyond mass, temperature and entropy (in this case volume and pressure) makes the derivation more cumbersome and less clear. We first recap the uncertainty principle based reasoning for the sake of comparison with our method (subsection~\ref{GUP}). In subsection~\ref{area}, we carry out a derivation of modified black hole temperature and entropy from minimal area. Lastly, subsection~\ref{estimates} introduces constraints on the minimal area and length implied by black hole evaporation.

Before, setting up the model and starting to play with it, we ought to say a few words about its limitations. Both the uncertainty principle and the minimal area method can determine temperature of the photons of Hawking radiation (up to a numerical factor). However, they do not say anything about their existence. To use them, we must \textit{assume} that black hole does emit radiation corresponding to a well-defined temperature. One can rigorously show that this indeed occurs and specify the necessary conditions~\cite{Hawking:1975,Visser:2003} (of course, the Hawking temperature is then found as a simple by-product of the calculations). However, to keep in tune with the rest of the paper, we will just support the existence of black hole radiation by a well known intuitive description introduced in the Hawking's seminal paper~\cite{Hawking:1975}. Consider pairs of virtual particles created just inside the horizon. Then, there exists a probability that the positive energy particle will tunnel outside of the horizon as Hawking radiation, while the negative energy one will be absorbed by the black hole, reducing its mass. Alternatively, one might consider a virtual particle pair just outside the horizon, with the negative energy particle tunnelling inside the black hole (while this model may sound somewhat hand-wavy, it can actually be used to derive many features of the Hawking radiation~\cite{Parikh:2000,Vanzo:2011}).

\subsection{Hawking temperature from uncertainty principle}
\label{GUP}

We begin by reviewing a heuristic derivation of Hawking temperature and modifications to it from the uncertainty principle~\cite{Scardigli:1995,Adler:2001,Alonso:2018a}. It will serve to provide a comparison with the minimal area approach we will introduce later on.

Consider a Schwarzschild black hole of mass $M$. A photon emitted from the black hole has uncertainty in its position comparable with the black hole's Schwarzschild radius, $r_S=2GM/c^2$~\cite{Kraus:1995}. The Heisenberg uncertainty principle (HUP) then implies a minimal uncertainty in photon's momentum
\begin{equation}
\Delta p=\frac{\hbar}{2\Delta x}\approx\frac{\hbar c^2}{4GM}.
\end{equation}
If we take $\Delta E=c\Delta p$ to be the typical energy of an emitted photon, the temperature of the radiation obeys
\begin{equation}
T\approx \frac{\Delta E}{k_B}\approx \frac{\hbar c^3}{4k_BGM},
\end{equation}
which agrees with the Hawking result for black hole temperature up to $1/2\pi$. However, since the above presented argument is only qualitative, some discrepancy in numerical factor can be expected. Therefore, the result must be corrected by a calibration factor of $1/2\pi$. Then, one recovers the Hawking temperature of a Schwarzschild black hole
\begin{equation}
T_H=\frac{\hbar c^3}{8\pi k_BGM}.
\end{equation}
To obtain a formula for the black hole entropy, one simply needs to integrate the equilibrium Clausius relation, $\text{d}S=c^2\text{d}M/T$, finding
\begin{equation}
S_B=4\pi k_B\frac{GM^2}{\hbar c}=k_B\frac{\mathcal{A}}{4l_P^2},
\end{equation}
where $\mathcal{A}$ denotes the horizon area. This is of course the well known expression for the Bekenstein entropy of a Schwarzschild black hole.

Let us remark that HUP sets a lower limit on the product of both uncertainties, but allows in principle arbitrarily precise determination of either position or momentum. However, thought experiments combining quantum mechanics with (even Newtonian) gravity indicate existence of a minimal resolvable length~\cite{Garay:1994en,Hossenfelder:2013}. Moreover, minimal length also arises in string theory~\cite{Strominger:1991}. To study its implications, one can introduce a modification of HUP, the generalised uncertainty principle (GUP). It reads
\begin{equation}
\Delta x\Delta p\ge\frac{\hbar}{2}\left(1+\alpha_0\frac{l_P^2}{\hbar^2}\Delta p^2\right),
\end{equation}
where $l_P=\sqrt{G\hbar/c^3}$ is the Planck length and $\alpha_0$ is a model dependent real number generally expected to be of the order of unity. The corresponding minimal length equals $l_{min}=\sqrt{\alpha_0}l_P$. There also exist variants of GUP incorporating minimal and/or maximal momentum~\cite{Iorio:2019}. However, we will limit our study to the above stated version, as it is represents the simplest modification of HUP necessary to incorporate minimal length and has the advantage of being supported by a variety of arguments, from simple thought experiments to more sophisticated calculations in various approaches to quantum gravity~\cite{Garay:1994en,Hossenfelder:2013}. Black hole temperature heuristically obtained from GUP then contains quantum gravity corrections coming from the existence of minimal length~\cite{Adler:2001}.

The previously described derivation, with HUP replaced by GUP, yields modified formulas for black hole temperature,
\begin{align}
\nonumber T_{GUP}=&\frac{2}{\frac{16\pi^2\alpha_0k_B^2l_P^2}{\hbar^2c^2}T_H}\left(1-\sqrt{1-\frac{16\pi^2\alpha_0k_B^2l_P^2}{\hbar^2c^2}T_H^2}\right) \\ =&T_H\left(1+\frac{4\pi^2\alpha_0k_B^2l_P^2}{\hbar^2c^2}T_H^2+O\left(\frac{\alpha_0^2k_B^4l_P^4}{\hbar^4c^4}T_H^4\right)\right),
\end{align}
and entropy,
\begin{equation}
S_{GUP}=\frac{k_B\mathcal{A}}{4l_{P}^2}-\frac{\pi\alpha_0k_B}{4}\ln \frac{\mathcal{A}}{\mathcal{A}_0} + O\left(\frac{k_B\alpha_0^2l_{P}^2}{\mathcal{A}}\right).
\end{equation}
This form of modified temperature has not been (to our best knowledge) confirmed by any rigorous method. However, a logarithmic correction term in black hole entropy was reported in many approaches to quantum gravity, including LQG~\cite{Kaul:2000,Meissner:2004}, string theory~\cite{Sen:2013} and AdS/CFT correspondence~\cite{Faulkner:2013}.

\subsection{Hawking temperature from minimal area}
\label{area}

We have seen how modified black hole temperature and entropy arise from the existence of minimal length. However, that turns out to be a slightly too strong assumption, as we show how to accomplish similar results by considering just a minimal area which, as we discussed in the introduction, represents a more general concept. Relaxing the requirement of minimal length is especially advantageous since some approaches to quantum gravity do not assume it, while still including a notion of minimal area.

One of the main candidate theories of quantum gravity, LQG, obtains a positive minimal eigenvalue of the area operator~\cite{Ashtekar:2021}. When considering loop quantum cosmology (LQC) within the improved dynamics prescription, this eigenvalue is identified with the ``area gap'', which is then treated as a physical minimal resolvable area given by $\Delta l_P^2=4\sqrt{3}\pi\gamma l_P^2$, with $\gamma$ being the Barbero-Immirzi parameter~\cite{Ashtekar:2006,Ashtekar:2009,Martin-Benito:2009,Rovelli:2014,Agullo:2016}. (Note, also, that the appearance of physical minimal area in LQG and LQC has been questioned in some works~\cite{Dittrich:2007,Bojowald:2020}). In an opposite way, the length operator in LQG does not provide a similar notion of minimal resolvable length~\cite{Bianchi:2008,Ashtekar:2021}. This framework then does not imply GUP in any straightforward way. Nevertheless, we do know that LQG leads to negative logarithmic correction to Bekenstein entropy~\cite{Kaul:2000,Meissner:2004} consistent with that implied by GUP. Generalising the derivation of modified entropy reviewed in the previous subsection to work just with minimal area could shed some light on the consistency of the phenomenological results.
	
Another approach to quantum gravity in which the difference of minimal length and minimal area becomes relevant is the proposal to quantise the area of black hole horizon~\cite{Bekenstein:1995,Bekenstein:1974,Bekenstein:1997,Medved:2008,Cardoso:2019,Agullo:2021}. It assumes that quantised area of a Kerr-Newman black hole has an evenly spaced spectrum, thus it directly introduces a minimal change in black hole area~\cite{Bekenstein:1995}. However, while the area is discretised, no specific realisation of a minimal area surface is envisioned. In other words, one does not divide the event horizon into some ``minimal area patches'' of a given shape, and the existence of minimal area only manifests in discrete (rather than continuous) changes of the horizon area. Hence, no constraints whatsoever are put on the minimal length. Since this idea offers a very simple model for quantum properties of black holes, its implications for their temperature and, especially, entropy are of interest, but they do not follow from the GUP-based argument.

Suppose there exists a minimal resolvable area, $\mathcal{A}_{min}$, implied by quantum gravity effects. Then, emission of a single photon of energy $c^2\delta M$ decreases the area of a Schwarzschild black hole event horizon at least by $\mathcal{A}_{min}$. If black hole's initial mass equals $M$, we have
\begin{align}
\nonumber \mathcal{A}_M-\mathcal{A}_{min}\ge\frac{16\pi G^2}{c^4}\left(M-\delta M\right)^2,\\
\delta M^2-2M\delta M+\frac{c^4}{16\pi G^2}\mathcal{A}_{min}\le 0 \label{dM},
\end{align}
where $\mathcal{A}_{M}=16\pi G^2M^2/c^4$ denotes the initial horizon area. Taking the value $\delta M c^2$, that saturates this inequality, as the typical energy of emitted photons, we get for their temperature
\begin{equation}
T\approx\frac{\delta Mc^2}{k_B}.
\end{equation}
We first apply this formula to a black hole large enough to satisfy $M>>\delta M$. In this approximation, we find $\delta M=c^4\mathcal{A}_{min}/32\pi G^2M$ and, therefore,
\begin{equation}
T\approx\frac{c^6\mathcal{A}_{min}}{32\pi k_BG^2M}.
\end{equation}
We can simplify the expression by writing $\mathcal{A}_{min}$ in terms of squared Planck length, $\mathcal{A}_{min}=\Delta l_P^2$, where $\Delta$ is some positive real number. Note that this entails no assumptions about the size of the minimal area as we keep $\Delta$ arbitrary. We reserve the discussion of its value for subsection~\ref{estimates}. Then, our expression for temperature reads
\begin{equation}
T\approx \frac{\Delta}{4}T_H.
\end{equation}
Due to a qualitative nature of our argument which may ignore some numerical factors, we must again introduce a calibration factor, in this case $4/\Delta$, to obtain the correct Hawking temperature. This factor explicitly depends on the size of the minimal area and for the typically considered values of $\Delta$, it turns out to be close to one (for instance, $\Delta\approx5.17$ often assumed in LQC~\cite{Ashtekar:2006} yields calibration factor approximately $1.29$). In conclusion, the photon temperature obeys
\begin{equation}
T=\frac{4}{\Delta}\frac{\delta Mc^2}{k_B}\approx\frac{\hbar c^3}{8\pi k_BGM},
\end{equation}
from which we easily find the Bekenstein entropy using the Clausius relation, just like in previous subsection. That is, the minimal area derivation recovers the standard expressions.

Upon deriving the standard formulae for black hole temperature and entropy, we turn to phenomenological quantum gravity corrections to them. To do so, we simply have to abandon the limit $M\gg \delta M$ (minimal area is already a consequence of quantum gravity, so no additional concept is required). Assuming again that the change in horizon area corresponds to $\mathcal{A}_{min}$ and solving the quadratic equation for $\delta M$ yields
\begin{equation}
\delta M=M\pm\sqrt{M^2-\frac{\Delta m_P^2}{16\pi}},
\end{equation}
where $m_P=\sqrt{\hbar c/G}$ is the Planck mass. We choose the minus sign as the plus sign gives a clearly unphysical solution for which $\delta M>M$, i.e., the emitted photon would have greater energy than the black hole itself (a similar ambiguity stemming from the existence of two solutions of a quadratic equation is also present in the GUP approach~\cite{Adler:2001}). Temperature corresponding to $\delta M=M-\sqrt{M^2-\Delta m_P^2/16\pi}$ equals
\begin{align}
\nonumber T_{mod}=&\frac{1}{2\pi\Delta\frac{k_B^2l_P^2}{\hbar^2c^2}T_H}\frac{Mc^2}{k_B}\left(1-\sqrt{1-4\pi\Delta\frac{k_B^2l_P^2}{\hbar^2c^2}T_H^2}\right) \\
=&T_H\left(1+\frac{\pi\Delta k_B^2G}{\hbar c^5}T_H^2+O\left(\frac{\Delta^2k_B^4G^2}{\hbar^2c^{10}}T_H^4\right)\right).
\end{align}
Once again, we find the entropy by integrating the Clausius relation. It yields
\begin{equation}
S_{mod}=\frac{k_B\mathcal{A}}{4l_{P}^2}-\frac{\Delta k_B}{16}\ln\left(\frac{\mathcal{A}}{\Delta l_P^2}\right)+O\left(\frac{k_B\Delta^2l_P^2}{\mathcal{A}}\right),
\end{equation}
where we chose $\mathcal{A}_{min}=\Delta l_P^2$ as a natural lower bound for the integration.

The previous formulae are implied only by the existence of a minimal area, without the necessity to introduce a minimal length. Nevertheless, one can consider a special case in which nonzero $\mathcal{A}_{min}$ indeed arises due to a minimal resolvable length, $l_{min}=\sqrt{\alpha_0}l_P$. While it clearly holds that $\mathcal{A}_{min}\propto l_{min}^2$, we are aware of no preferred exact relation between both quantities. However, one of the fairly natural options is to choose $\mathcal{A}_{min}$ as an area of a 2-sphere whose radius equals $l_{min}$, i.e., $\mathcal{A}_{min}=4\pi l_{min}^2=4\pi\alpha_0 l_P^2$. Then, the modified temperature and entropy formulas become
\begin{align}
\nonumber T_{GUP}=&\frac{2}{\frac{16\pi^2\alpha_0k_B^2l_P^2}{\hbar^2c^2}T_H}\left(1-\sqrt{1-\frac{16\pi^2\alpha_0k_B^2l_P^2}{\hbar^2c^2}T_H^2}\right) \\
=&T_H\left(1+\frac{4\pi^2\alpha_0k_B^2l_P^2}{\hbar^2c^2}T_H^2+O\left(\frac{\alpha_0^2k_B^4l_P^4}{\hbar^4c^4}T_H^4\right)\right),\\
S_{GUP}=&\frac{k_B\mathcal{A}}{4l_{P}^2}-\frac{\pi\alpha_0k_B}{4}\ln \frac{\mathcal{A}}{\mathcal{A}_0} + O\left(\frac{k_B\alpha_0^2l_{P}^2}{\mathcal{A}}\right),
\end{align}
in precise agreement with results obtained from GUP. If one considers a different relation between $\mathcal{A}_{min}$ and $l_{min}$, the numerical factors in the correction terms differ, but basic structure of the expressions remains the same.

\subsection{Estimation of free parameters}
\label{estimates}

All the formulae for modified black hole temperature and entropy we discussed contain a single undetermined model dependent parameter, either $\alpha_0$, or $\Delta$. (Recall that $\alpha_0$ is present in the statement of GUP and we introduced $\Delta$  as a measure of minimal area). While both are widely believed to be of the order of unity based on the theoretical frameworks in which they arise, the current experimental constraints are far less stringent. The lowest reported upper limit on $\alpha_0$ we are aware of (coming from measurements of frequency shifts of harmonic oscillators) is $\alpha_0\lesssim10^{6}$~\cite{Bushev:2019} and most methods yield much higher upper bounds~\cite{Gao:2017,Girdhar:2020}. Here, upon adding a new game piece, we are able to introduce a novel theoretical estimate providing a fairly stringent bound, $\alpha_0\approx\Delta\lesssim10^1$. 

The necessary additional game piece is the black body character of Hawking radiation. The average entropy per photon of black body radiation equals~\cite{Alonso:2018b}
\begin{equation}
\langle\hat{S}\rangle=\frac{\pi^4}{30\zeta\left(3\right)}k_B\approx2.70k_B.
\end{equation}
After including the standard deviation, we get \mbox{$\langle\hat{S}\rangle\approx\left(2.70\pm1.75\right)k_B$}. Hawking radiation has a black body spectrum at least up to the late stages of evaporation, when quantum gravity effects might modify it. Since the average Clausius entropy per photon of black body radiation of any origin equals $\langle\hat{S}\rangle$, this result also holds for Hawking radiation at the level of macroscopic semiclassical thermodynamics~\cite{Alonso:2018b}. According to generalised second law of thermodynamics~\cite{Bekenstein:1973}, the average change of Bekenstein entropy due to emission of one photon, $\langle\Delta S_B\rangle$, obeys
\begin{equation}
\langle\hat{S}\rangle+\langle\Delta S_B\rangle\ge0.
\end{equation}
This means that the average decrease of Bekenstein entropy per emitted photon is \textit{at most} \mbox{$\vert\langle\Delta S_B\rangle\vert\approx\left(2.70\pm1.75\right)k_B$}.  Now, combine this bound with the existence of minimal area, that implies the decrease of Bekenstein entropy per emitted photon is \textit{at least} $\left(\Delta/4\right)k_B$. Comparing both limits directly gives us an upper bound on $\Delta$, specifically $\Delta\lesssim10.80$. This value fits very well with the common expectation $\Delta\approx1$.

We have seen that in any minimal length scenario the choice $\Delta=4\pi\alpha_0$ leads to an equivalence between corrections derived from GUP and from minimal area. Furthermore, such a choice is geometrically well motivated as an area of a 2-sphere whose radius is equal to the minimal length. Thus, any estimate of the minimal area also directly gives us the minimal length, provided that such a length exists in the model of quantum gravity we consider. Of course, the precise relation $\Delta=4\pi\alpha_0$ is a bit arbitrary and not necessarily correct. Hence, we only argue that $\alpha_0$ (if it exists) is roughly comparable with $\Delta$ and any upper bound on $\Delta$ also represents an upper bound on $\alpha_0$ (while not excluding the option $\Delta>0$, $\alpha_0=0$, i.e., there is minimal area but no minimal length). Therefore, we are able to provide the general bound $\alpha_0\lesssim10.80$. Note that if we were to take seriously the relation $\Delta=4\pi\alpha_0$, we would have gotten even more stringent bound $\alpha_0\lesssim0.86$.

Let us recall some theoretical predictions for $\Delta$. First, if the area gap of LQC indeed represents a physical minimal area, it implies $\Delta\approx5.17$. This value comes from fixing the Barbero-Immirzi parameter by demanding that one recovers Bekenstein entropy for black holes~\cite{Ashtekar:2006,Meissner:2004}. Second, in the context of quantisation of black hole horizon area, the value $\Delta=4\ln2\approx2.77$ has been suggested as the most natural one~\cite{Bekenstein:1995}. It is chosen so that the minimal change of (dimensionless) black hole entropy is precisely $1$ bit. We can see that not only both values fit within our predicted upper limit $\Delta\lesssim10.80$, but both are rather close to it, implying that the bound we found is already quite restrictive.

Actually, we can make a more ambitious attempt and not only constrain $\Delta$ from above, but provide an estimation of its value. The price to pay is a loss of robustness, so we may more appropriately consider it a guess for an approximate value than a genuine prediction. To do so, we associate the minimal entropy per photon with the average entropy minus its standard deviation, i.e., $\Delta/4\approx2.70-1.95$ (one can still expect a significant number of photons with this entropy, so it should not be smaller than its minimal value). In this way, we find \mbox{$\Delta\approx3.80$}, in a very good agreement with other theoretical predictions.

Let us stress that the argument for the upper value of $\mathcal{A}_{min}$ does not depend on the heuristic reasoning we discussed in previous subsections. Only two assumptions are required; that a minimal resolvable area exists (otherwise, estimating it feels rather pointless) and that average entropy per photon of Hawking radiation is about $2.70k_B$, at least in the early stages of evaporation, when quantum gravity effects can be safely neglected (while it was proposed that the discreteness of area causes the Hawking radiation to have a line spectrum instead of a continuous one, this should not significantly affect the entropy per photon~\cite{Bekenstein:1995,Bekenstein:1997}).

\section{Thermodynamics of causal diamonds}
\label{GLCD}

\begin{figure}[tbp]
	\centering
	\includegraphics[width=8.6cm,origin=c,trim={2.0cm 5.0cm 20.0cm 5.0cm},clip]{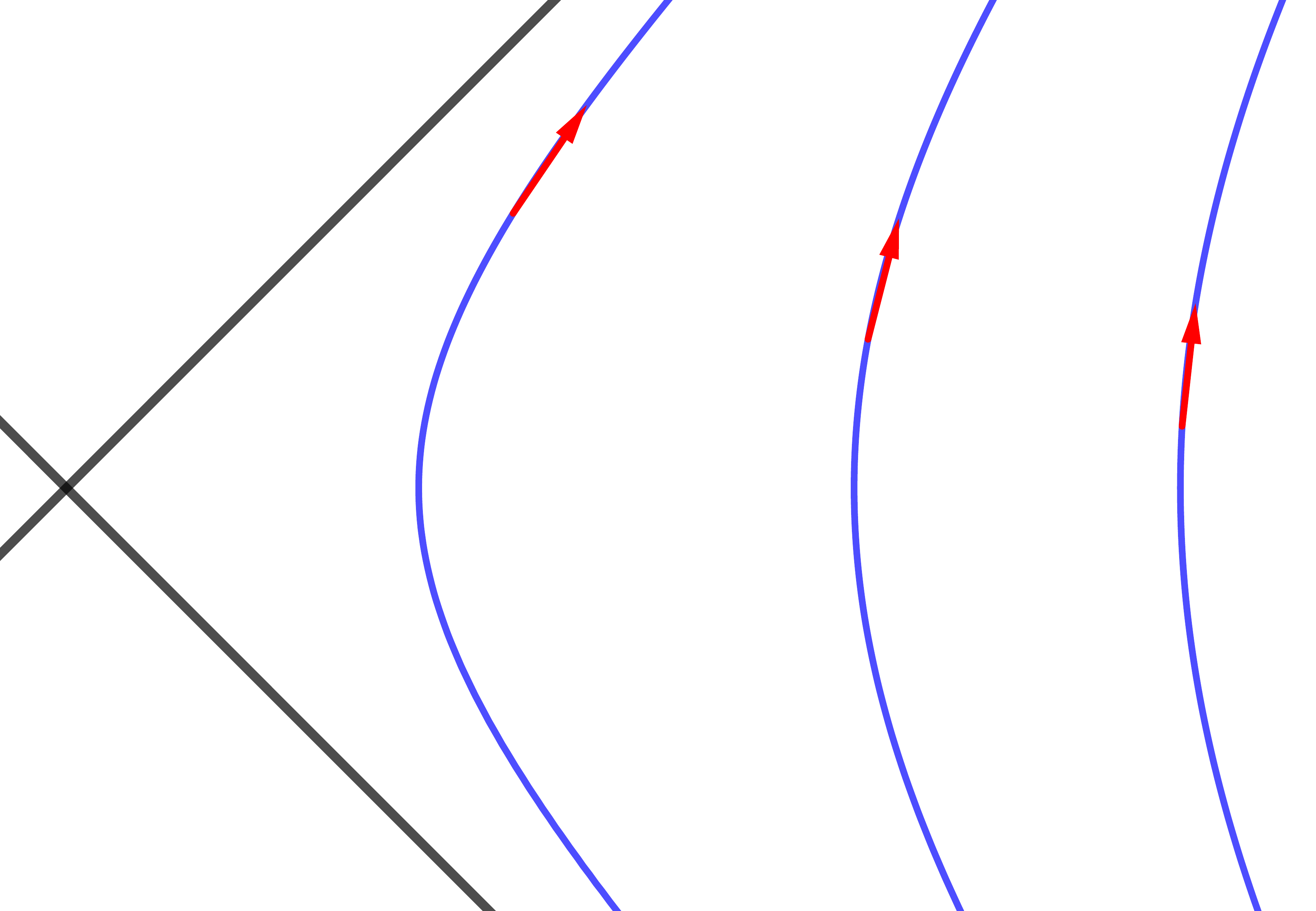}
	\caption{\label{Rindler} A simple scheme of a Rindler horizon. We depict only the right Rindler wedge in detail, the left one is symmetric to it. The horizon is represented by oblique lines, curved lines are wordlines of a few selected uniformly accelerating observers whose velocities are denoted by arrows. Even though the observers have different accelerations and the temperatures they measure thus differ, all of them have access to the same region of spacetime (the right Rindler wedge) and perceive the same entanglement entropy.}
\end{figure}

It has been shown in the literature that heuristic derivations from HUP work both for the Hawking temperature of Schwarzschild black holes and for the Unruh temperature measured by uniformly accelerating observers~\cite{Scardigli:1995}. From them, on one hand, Hawking temperature and equilibrium Clausius relation provide a more or less direct route to obtain an expression for black hole entropy. On the other hand, while one can assign entanglement entropy to a horizon perceived by an accelerating observer who measures non-zero Unruh temperature~\cite{Solodukhin:2011}, there appears to be no way to obtain an expression for this entropy from the temperature. If this were possible, it would establish a heuristic analogy between thermodynamics of black holes and observer-dependent horizons. However, one cannot connect entanglement entropy and Unruh temperature for horizons perceived by eternal, uniformly accelerating observers (Rindler horizons, sketched in figure~\ref{Rindler}), because a single such horizon corresponds to a class of observers with different accelerations and, consequently, different temperatures.

In this section, we will argue that a connection between both quantities, Unruh temperature and entropy, exists in the case of causal diamonds. We can understand these structures as a convenient way to define small regions filling the spacetime. Relating their temperature and entropy is possible because causal diamonds posses a ``preferred'' Unruh temperature; the one measured by finite lifetime inertial observers~\cite{Martinetti:2003,Arzano:2020}.

\subsection{Geodesic local causal diamonds}

Let us begin by introducing the structure that will allow us to obtain a connection between Unruh temperature and entropy: geodesic local causal diamonds (GLCD). In this subsection we will briefly explain their construction and relevant properties before starting to play with them. One can find more detailed descriptions of these objects, e.g. in~\cite{Gibbons:2007,Jacobson:2017,Wang:2019,Jacobson:2019}.

\begin{figure}[tbp]
	\centering
	\includegraphics[width=8.6cm,origin=c,trim={0.1cm 2.4cm 36.7cm 1.5cm},clip]{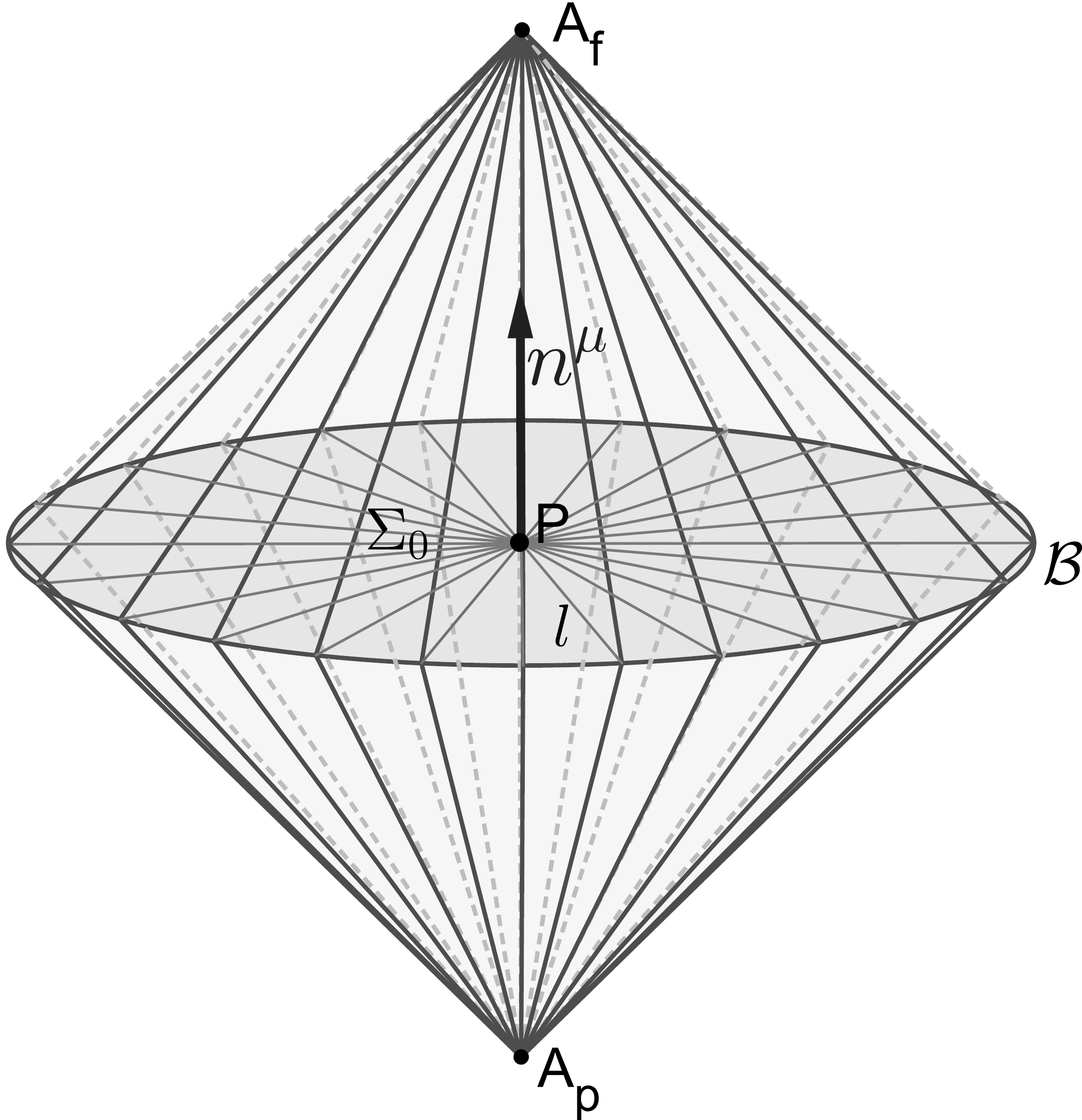}
	\caption{\label{diamond} A GLCD with the origin in point $P$ (the angular coordinate $\theta$ is suppressed). $\Sigma_0$ denotes a spatial geodesic ball of radius $l$ (some of the geodesics forming it are represented as grey lines), whose boundary is an approximate 2-sphere $\mathcal{B}$. The normal to $\Sigma_0$ is the timelike vector $n^{\mu}$. The tilted lines starting at the past apex $A_p$ ($t=-l/c$) and going to the future apex $A_f$ ($t=l/c$) represent some of the null geodesic generators of the GLCD boundary. Geodesic ball $\Sigma_0$ is the spatial cross-section of the future domain of dependence of $A_p$ and the past domain of dependence of $A_f$ at $t=0$.}
\end{figure}

Choose any spacetime point $P$ and an arbitrary unit timelike vector $n(P)$. In every direction orthogonal to $n$ send out of $P$ geodesics of parameter length $l$ to form a geodesic 3-ball, $\Sigma_0$. The region of spacetime causally determined by $\Sigma_0$ is known as a GLCD. The construction of this object is illustrated in figure~\ref{diamond} for the sake of clarity. The boundary, $\mathcal{B}$, of $\Sigma$ is approximately a 2-sphere whose area equals~\cite{Jacobson:2015}
\begin{equation}
\mathcal{A}=4\pi l^2-\frac{4\pi}{9}l^4G_{00}\left(P\right)+O\left(l^5\right),
\end{equation}
where $G_{00}=G_{\mu\nu}n^{\mu}n^{\nu}$. The boundary $\mathcal{B}$, understood as a 2-surface embedded in a spatial 3-surface containing $\Sigma_0$, has extrinsic curvature~\cite{Jacobson:2019}
\begin{equation}
k=\frac{2}{l}.
\end{equation}
The GLCD is endowed with an approximate (up to $O\left(l^3\right)$ curvature dependent terms) conformal Killing vector~\cite{Jacobson:2015}
\begin{equation}
\zeta=C\left(\left(l^2-t^2-r^2\right)\frac{\partial}{\partial t}-2rt\frac{\partial}{\partial r}\right),
\end{equation}
where $C$ represents an arbitrary normalisation constant. It is often set to $C=1/2l$ so that $\zeta$ has a unit surface gravity~\cite{Jacobson:2015}. However, we will keep $C$ unspecified and consider arbitrary surface gravity denoted by $\kappa$ to clearly demonstrate that the value of $C$ is irrelevant for our discussion. From the definition of $\zeta$ one easily sees that the null boundary of the GLCD forms a conformal Killing horizon.

\begin{figure}[tbp]
	\centering
	\includegraphics[width=8.6cm,origin=c,trim={5.5cm 0.1cm 6.3cm 0.1cm},clip]{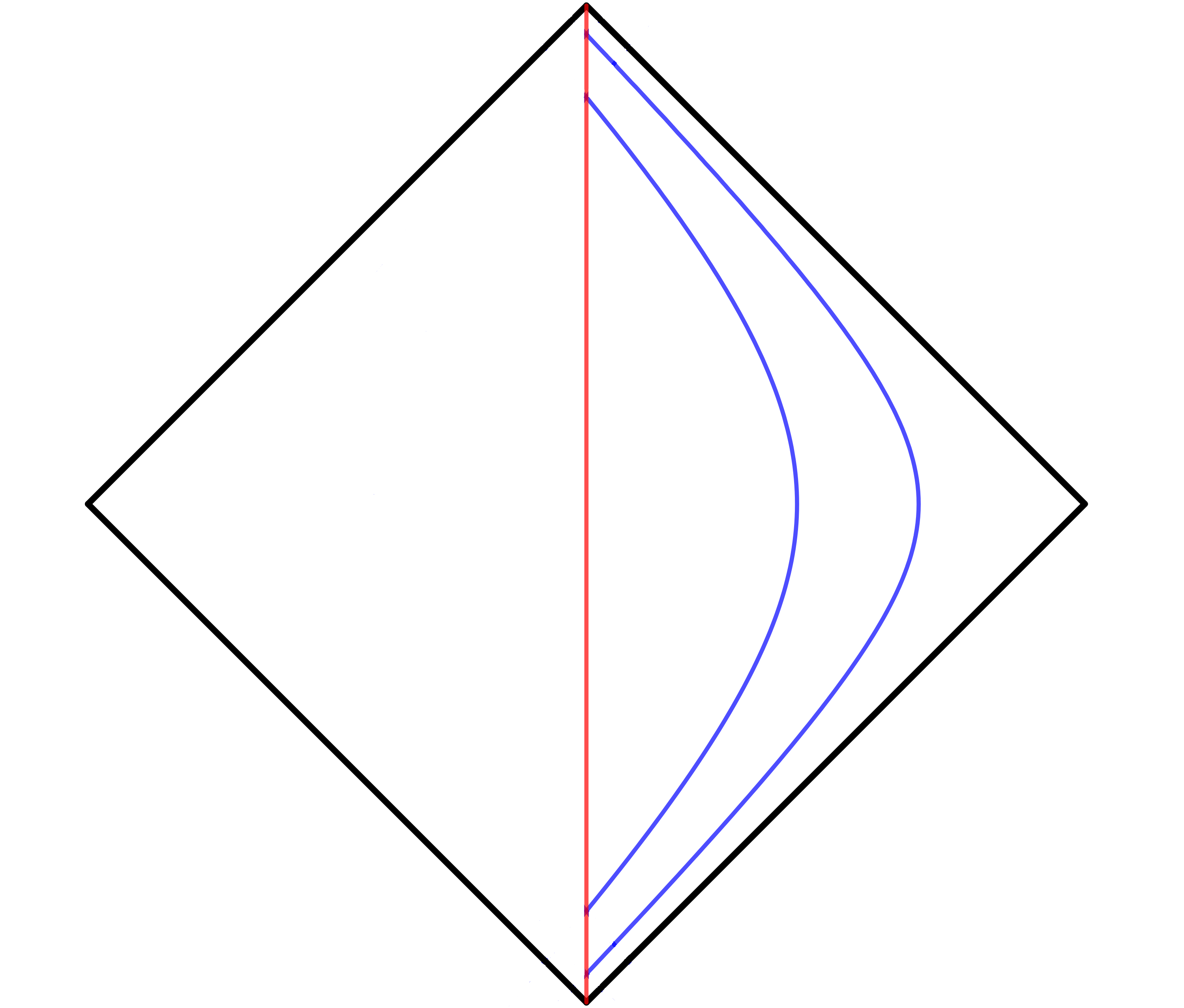}
	\caption{\label{observers} Observers moving inside a causal diamond who perceive nonzero Unruh temperature. Oblique lines form the diamond's conformal Killing horizon. The vertical line represents the inertial finite lifetime observer, whose existence starts in the past apex of the diamond and ends in its future apex, corresponding to lifetime $2l/c$. The curved lines are example worldlines of observers who travel inside the diamond with an acceleration of constant magnitude $a$.}
\end{figure}

In order to define temperature associated with a GLCD, let us point out the appearance of two different concepts of it in the literature. One option refers to the conformal Killing vector $\zeta$, defining the Hawking temperature of the corresponding horizon as \mbox{$T_{H}=\hbar\kappa/2\pi k_Bc$}~\cite{Jacobson:2019}. Due to the arbitrary normalisation constant $C$ in the definition of $\zeta$, temperature $T_{H}$ can take any value. The second possibility is considering the Unruh temperature measured by accelerating observers moving inside the GLCD. While these observers are not infinitely uniformly accelerating, they will approximately measure the usual Unruh temperature, $T_{U}=\hbar a/2\pi k_Bc$, as long as the magnitude of their acceleration is constant and satisfies $a\gg c^2/l$~\cite{Barbado:2012}. Furthermore, the isometry between a causal diamond in flat spacetime and a Rindler wedge allows one to find the Unruh temperature even for small accelerations~\cite{Martinetti:2003}. In particular, even a finite lifetime inertial observer travelling between the apices of the diamond (see figure~\ref{observers}) measures finite Unruh temperature~$T_{inertial}=\hbar c/2k_Bl$~\cite{Martinetti:2003,Arzano:2020}. In the following, we will concentrate on recovering $T_{inertial}$ from the existence of minimal area.

To find the temperature and entropy of causal diamonds from minimal area, we still need one more game piece, some relation for small variations of causal diamond's area similar to equation~(\ref{dM}) for a Schwarzschild black hole. Thankfully, changes of area and volume of causal diamonds obey an equation analogous to the first law of black holes mechanics~\cite{Jacobson:2019},
\begin{equation}
\delta H_m=-\frac{c^4}{8\pi G}\kappa\delta\mathcal{A}+\frac{c^4}{8\pi G}\kappa k\delta V,
\end{equation}
where $\delta H_m$ denotes the change of the matter Hamiltonian. Note that the first law of causal diamond can be rigorously proved using the Noether charge formalism, without making any reference to thermodynamics~\cite{Jacobson:2019}. Therefore, we can use it to heuristically obtain the expressions for diamond's temperature and entropy without making a circular argument.

\subsection{Diamond temperature and entropy from minimal area}	

For simplicity, we will focus on a GLCD that is initially in Minkowski spacetime. Generalisation to a sufficiently small GLCD in an arbitrary spacetime would be straightforward, but needlessly messy. Let us consider a photon being detected by a finite lifetime inertial observer associated with the GLCD. The presence of the photon will give rise to a small spacetime curvature, decreasing the area of the GLCD\footnote{Treating the influence of the Unruh effect on spacetime geometry in this way is admittedly rather cavalier. However, we believe it suffices for the heuristic argument we intend to make.}. For the variation of matter Hamiltonian it holds $\delta H_m=4\pi l^4\kappa\delta T_{00}/15$~\cite{Jacobson:2019} (the same result applies even for quantised conformal fields, just with $\delta T_{00}$ replaced by the expectation value, $\delta\langle T_{00}\rangle$~\cite{Jacobson:2015}). The first law of causal diamonds then implies
\begin{equation}
\frac{4\pi}{15}l^4\kappa\delta T_{00}=-\frac{c^4}{8\pi G}\kappa\delta\mathcal{A}+\frac{c^4}{8\pi G}\kappa k\delta V.
\end{equation}
We stress that $\kappa$ drops out of the equation and, therefore, the normalisation of $\zeta$ indeed does not affect our reasoning. Furthermore, variations of volume and area are not independent. Since we keep $l$ constant, it holds $\delta V=l\delta\mathcal{A}/5$~\cite{Jacobson:2015}. If we assume that $\delta\mathcal{A}$ corresponds to the minimal area, $\mathcal{A}_{min}$, and identify the photon's energy as $\delta E=V\delta T_{00}=4\pi l^3\delta T_{00}/3$ (i.e., energy density times volume), we find
\begin{equation} \label{1lawcd}
\frac{1}{5}l\delta E=\frac{3}{5}\frac{c^4}{8\pi G}\mathcal{A}_{min}.
\end{equation}
Therefore, temperature measured by the inertial finite lifetime observer satisfies
\begin{equation}
T_{inertial}\approx\frac{\delta E}{k_B}=\frac{1}{k_B}\frac{3c^4\mathcal{A}_{min}}{8\pi G}\frac{1}{l}=\frac{3\Delta}{8}\frac{\hbar c}{\pi k_B}\frac{1}{l}.
\end{equation}
Here one can see that temperature is proportional to $1/l$ in accordance with the result obtained from isometry between a causal diamond and a Rindler wedge~\cite{Martinetti:2003,Arzano:2020}. However, to get the exact formula for Unruh temperature, we must again add a calibration factor, $8/3\Delta$.\footnote{The necessary factor differs from the value calibrating the case of the Schwarzschild black hole, $4/\Delta$. This difference is expected as the processes we study in both cases are somewhat different. For a black hole, we consider change in its Schwarzschild radius and, consequently, area due to the emission of a photon. However, length scale $l$ associated with a causal diamond is held fixed. Instead, the change in area is brought about by the backreaction of the spacetime geometry to a photon of Unruh radiation.} Multiplying the expression for $T_{inertial}$ by the calibration factor, we straightforwardly obtain
\begin{equation}
T_{inertial}=\frac{\hbar c}{\pi k_B}\frac{1}{l}=\frac{\hbar c}{2\pi k_B}k.
\end{equation}
From that, finding the entropy of the GLCD's horizon is somewhat more difficult than in the case of a Schwarzschild black hole, as the first law of causal diamonds involves an extra term corresponding to the variation of volume. To identify a notion of entropy, we interpret the quantities present in the previously defined first law of causal diamonds in the standard thermodynamic form of
\begin{equation}
\text{d}U=T\text{d}S+p\text{d}V.
\end{equation}
Thus, if we want to treat the left-hand-side of equation~\eqref{1lawcd} as heat, it must have dimensions of energy. However, we have seen that it is actually proportional to $l\delta E$. To amend this, we can multiply the entire equation by some quantity with dimensions of inverse length. The only such natural quantity associated with the GLCD is extrinsic curvature, $k=2/l$ (another option would be $1/l_P$, but since we do not consider any minimal length scenario, it appears to be a rather artificial choice). Then, the first law yields
\begin{align}
\nonumber \frac{2}{5}\delta E=&-\frac{c^4}{8\pi G}k\delta\mathcal{A}+\frac{c^4}{8\pi G}k^2\delta V \\
\nonumber =&-T_{inertial}\frac{k_Bc^3}{4G\hbar}\delta\mathcal{A}+\frac{c^4}{8\pi G}k^2\delta V \\
\equiv& T_{inertial}\delta S+p\delta V,
\end{align}
where we have identified $p=c^4k^2/8\pi G$, that has the correct dimensions of pressure, $kg\cdot m^{-1}\cdot s^{-2}$. 

Now we can consider a simultaneous change of area and volume with the matter content held fixed, obtaining
\begin{equation}
\text{d}S=-\frac{p}{T_{inertial}}\text{d}V,
\end{equation}
and, therefore,
\begin{equation}
S=\frac{\pi k_Bc^3l^2}{G\hbar}=k_B\frac{\mathcal{A}}{4l_P^2}.
\end{equation}
We have arrived precisely at the Bekenstein entropy of the GLCD's conformal horizon, in accordance with previous assumptions in thermodynamics of spacetime\mbox{~\cite{Jacobson:2015,Jacobson:2019}}.

The most interesting feature of our result is the connection between the Unruh temperature and the entropy associated with an observer dependent object. The possibility of such a connection has been hinted at in previous papers dealing with the temperature of finite lifetime inertial observers~\cite{Martinetti:2003,Arzano:2020}. It occurs since a GLCD is, much like a Schwarzschild black hole, fully characterised by a single length scale, $l$, and it holds $ST_{inertial}=c^4l/G$. Similarly, one gets \mbox{$ST_{H}=c^4r_S/4G$} for the case of a Schwarzschild black hole. A connection of this kind does not exists between the Unruh temperature and entropy associated with a Rindler wedge. In other words, one can assign a preferred temperature to causal diamonds (the one measured by inertial observers), but not to Rindler wedges (where none of the accelerating observers is privileged, at least as far as physics is concerned).

\subsection{Modified temperature and entropy for causal diamonds}

Following our previous treatment of a Schwarzschild black hole, a natural question now is whether we can use the method developed in previous subsection to find phenomenological quantum gravity corrections to GLCD's temperature and entropy. The direct answer is that the process is rather tricky. Nevertheless, basic features of the modified expressions can be guessed fairly easily. To do so, we include the change of volume due to curvature to the left-hand-side of the first law and leave the right-hand-side as before
\begin{equation}
\frac{1}{5}l\delta E-\frac{27}{70\pi}\frac{1}{l}\mathcal{A}_{min}\delta E=\frac{3}{5}\frac{c^4}{8\pi G}\mathcal{A}_{min}.
\end{equation}
The modified temperature then equals
\begin{align}
\nonumber T_{mod}=&T_{inertial}\bigg(1+\frac{27\pi\Delta Gk_B^2}{14\hbar c^5}T_{inertial}^2 \\
&+O\left(\frac{\Delta^2G^2k_B^4}{\hbar^2c^{10}}T_{inertial}^4\right)\bigg),
\end{align}
and has the same structure as both modifications of the Hawking temperature implied by the minimal area and recently proposed modifications of the Unruh temperature due to GUP~\cite{Scardigli:2018jlm}. For the entropy, we obtain in the same way as previously (assuming unmodified pressure)
\begin{equation}
S_{mod}=k_B\frac{\mathcal{A}}{4l_P^2}-\frac{27\Delta}{28}k_B\ln\left(\frac{\mathcal{A}}{\mathcal{A}_{min}}\right)+O\left(\frac{k_B\mathcal{A}_{min}}{\mathcal{A}}\right).
\end{equation}
We can see that the result is qualitatively in agreement with the modified Bekenstein entropy. In this way, we have found that, up to a numerical factor, the existence of the minimal area implies the same modified entropy for a black hole and a causal diamond. Furthermore, it agrees with logarithmic corrections to the entanglement entropy of a 2-sphere in Minkowski spacetime~\cite{Solodukhin:2010}.

The previous procedure of course disregards higher order corrections to the right hand side of the first law. One can actually attempt a more precise calculation, approximating the effect of Unruh radiation on curvature in terms of spatially homogeneous, isotropic and flat metric. In this case, the first law yields
\begin{equation}
\frac{1}{5}l\delta E-\frac{27}{70\pi}\frac{8\pi G}{3c^4}\delta E^2=\frac{3}{5}\frac{c^4}{8\pi G}\mathcal{A}_{min}-\frac{88G}{7875c^4}\delta E^2,
\end{equation}
and, up to coefficients in correction terms, we reach the same conclusions. Anyway, we do know that we cannot find the exact numerical factors using our method and we expect that neither this correction nor any other more refined approach is going to provide new qualitative effects. Then, its analysis is not relevant for our purposes. The signs and orders of magnitude of the correction terms are already correctly captured by the simple estimate we made above, and would not be modified by further corrections.

To sum up, we found that the relation between temperature and entropy of a causal diamond is analogous to that between temperature and entropy of a Schwarzschild black hole not only semiclassically, but even when leading order quantum gravity corrections are taken into account. It would be interesting to rederive this relation in a more rigorous study and to find out how far we can extend the similarity of causal diamonds and black holes. We will address these questions in a future work.

\section{Discussion}
\label{discussion}

We have heuristically derived modified Hawking temperature and Bekenstein entropy of a Schwarzschild black hole from the existence of minimal area, generalising a similar derivation from minimal length. The modified entropy containing a term logarithmic in horizon area qualitatively agrees with results obtained, e.g. in LQG, AdS/CFT and calculations of entanglement entropy. Furthermore, we have used the known semiclassical properties of Hawking radiation to constrain the size of the minimal area. The upper bound we found, $\mathcal{A}_{min}\lesssim10.80l_P^2$, is of the same order as theoretical proposals for minimal area, but somewhat larger than them. 

We have also extended our heuristic derivation to causal diamonds, obtaining a formula for the Unruh temperature measured by finite lifetime inertial observers and for the entropy of diamond's horizon. Furthermore, we have proceeded to derive modifications of temperature and entropy due to low energy quantum gravity effects, finding results consistent with the proposal of GUP-modified Unruh temperature and logarithmic corrections to entanglement entropy of a 2-sphere, respectively.

Let us stress that the entropy modifications we have obtained for black holes and causal diamonds contain only microcanonical corrections, coming from more precise knowledge of microstates responsible for entropy due to insights from quantum gravity (phenomenologically captured in the existence of minimal length/area). A complete treatment of the logarithmic term in the entropy would require adding canonical corrections. These ones arise due to thermal fluctuations at fixed Hawking/Unruh temperature and increase the entropy. It has been argued that canonical correction to black hole entropy should be at least \mbox{$\Delta S_{c}\gtrsim\left(3k_B/2\right)\ln\left({\mathcal{A}/\mathcal{A}_{min}}\right)$}~\cite{Medved:2005}. Our upper bound on the minimal area, $\mathcal{A}_{min}\lesssim10.80l_P^2$, and heuristic formula for modified entropy together yield an upper bound for the magnitude of microcanonical corrections, \mbox{$\vert\Delta S_{m}\vert\lesssim0.68k_B\ln\left({\mathcal{A}/\mathcal{A}_{min}}\right)$}. Combination of both bounds would then imply \mbox{$\Delta S_{c}+\Delta S_{m}\gtrsim0.82k_B\ln\left({\mathcal{A}/\mathcal{A}_{min}}\right)>0$}, i.e., the total logarithmic correction to black hole entropy would be positive. Since the overall sign of the logarithmic term remains an open issue, with some implications for the final stages of black hole evaporation~\cite{Alonso:2021,Arzano:2005,Solodukhin:2020}, it would be of interest to explore this issue further in the future. Moreover, working out the complete logarithmic term in entropy of causal diamonds could help to constrain quantum phenomenological gravitational dynamics proposed by the authors of this paper~\cite{Alonso:2020b}.

Recently, a heuristic derivation of Bekenstein bound and its modification due to quantum gravity effects from uncertainty relations has been proposed~\cite{Scardigli:2021}. The authors have even derived the uncertainty relations from the Bekenstein bound. Finding a similar relationship between the Bekenstein bound and minimal area would strengthen the notion that minimal area has the same implications for black hole thermodynamics as minimal length. Moreover, it would connect two Bekenstein's ideas, the upper bound on entropy contained in a given region and the quantisation of area. This direction will be further explored in a future work.

Besides our results concerning black hole physics, we have also found a new relation between Unruh temperature and entropy of causal diamonds. This connection we have heuristically established strengthens the analogy between thermodynamics of causal diamonds and black holes. Both posses a ``preferred'' notion of temperature (black holes the Hawking temperature measured by inertial observers at infinity and causal diamonds the Unruh temperature measured by finite lifetime inertial observers), and entropy of both can be derived from temperature via the equilibrium Clausius relation. Next step will be trying to confirm the relation of entropy and temperature by more rigorous methods, especially since it could have implications for deriving gravitational dynamics from thermodynamics of causal diamonds\mbox{~\cite{Jacobson:2015,Bueno:2017,Svesko:2019,Alonso:2020a}}.

To conclude, let us remark that in this paper, we have introduced a new basic structure to the game of thermodynamics of spacetime and its connection to phenomenology of quantum gravity. Here, we do not pretend to establish solid results based on first principles, but instead explore new insights and relations that will be worth a more detailed treatment in future works.

\acknowledgments

The authors want to acknowledge Javier Olmedo for discussions on the minimal area appearance in the context of LQC. AA-S is supported by the ERC Advanced Grant No. 740209. M.L. is supported by the Charles University Grant Agency project No. GAUK 297721. This work is also partially supported by Project. No. MICINN PID2020-118159GB-C44.

\end{document}